\documentclass{article}

\usepackage{arxiv}

\usepackage[utf8]{inputenc} 
\usepackage[T1]{fontenc}    
\usepackage{hyperref}       
\usepackage{url}            
\usepackage{booktabs}       
\usepackage{amsfonts}       
\usepackage{amsmath}        
\usepackage{nicefrac}       
\usepackage{microtype}      
\usepackage{lipsum}		    
\usepackage{subfig}         
\usepackage{ragged2e}       
\usepackage{graphicx}
\usepackage{natbib}
\usepackage{doi}

\title{Implicitization of biquadratic Bézier triangle and quadrilateral surfaces}

\author{ \href{https://orcid.org/0000-0001-9033-6365}{\includegraphics[scale=0.06]{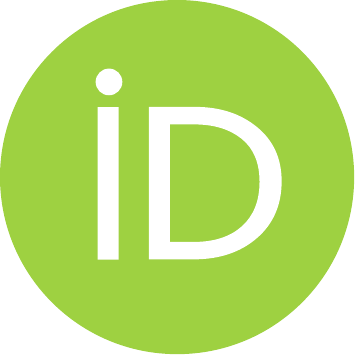}\hspace{1mm}Jackson Borchardt and Saul Kato} \\
	Weill Institute for Neurosciences, Department of Neurology\\
	University of California, San Francisco\\
	San Francisco, CA 94107 \\
	\texttt{jackson.borchardt2 -at- ucsf.edu | saul.kato -at -ucsf.edu}
}

\renewcommand{\shorttitle}{Implicitization of biquadratic surfaces}

\hypersetup{
pdftitle={Implicitization of biquadratic Bézier triangle and quadrilateral surfaces
},
pdfsubject={q-bio.NC, q-bio.QM},
pdfauthor={David S.~Hippocampus, Elias D.~Striatum},
pdfkeywords={First keyword, Second keyword, More},
}

\begin{document}
\maketitle
\begin{abstract}
We produce implicit equations for general biquadratic (order 2x2) Bézier triangle and quadrilateral surface patches and provide function evaluation code, using modern computing resources to exploit old algebraic construction techniques.
\end{abstract}


\section{Introduction}
In the fields of computer graphics, computer-aided design, and physical simulation, it is common to model complex curved surfaces as piecewise compositions of simpler surface patches. One type of parametric surface family used for such piecewise compositions is the Bézier surface, a two-dimensional extension of the Bézier curve, a polynomial curve parameterized by a fixed number (depending on the order of the surface) of control points in space.

For any parametrically defined surface, there exists an equivalent implicit equation representation $F(x,y,z)=0$, which has several utilities. An implicit equation can be used to directly evaluate whether an arbitrary point in space lies on, on one side, or on the other side of a given surface, which is useful for hit/collision detection in physical simulation and for voxelization of objects. An implicit equation can also be used to compute the intersection of one surface and another parametric surface by direct substitution, which is useful for ray tracing and computation of wireframes of objects built from unions of solid primitives. Implicitization procedures for polynomial surfaces have been studied for over a century; however, due to the unwieldy size of the resulting implicit equations, they have not been carried out for general cases of polynomial surfaces, even those of very low order. The advent of automated symbolic computation and the explosion of computational resources warrants a revisitation of algebraic elimination theory to produce useful closed-form implicit expressions of expressive mathematical surfaces for generative modeling applications.

We derive implicit equations for general biquadratic (order 2x2) Bézier surfaces with quadrilateral and triangular input domains and provide evaluation code in Python.

\section{Background}
\label{sec:background}
\begin{figure}
	\centering
    \includegraphics[height=2in]{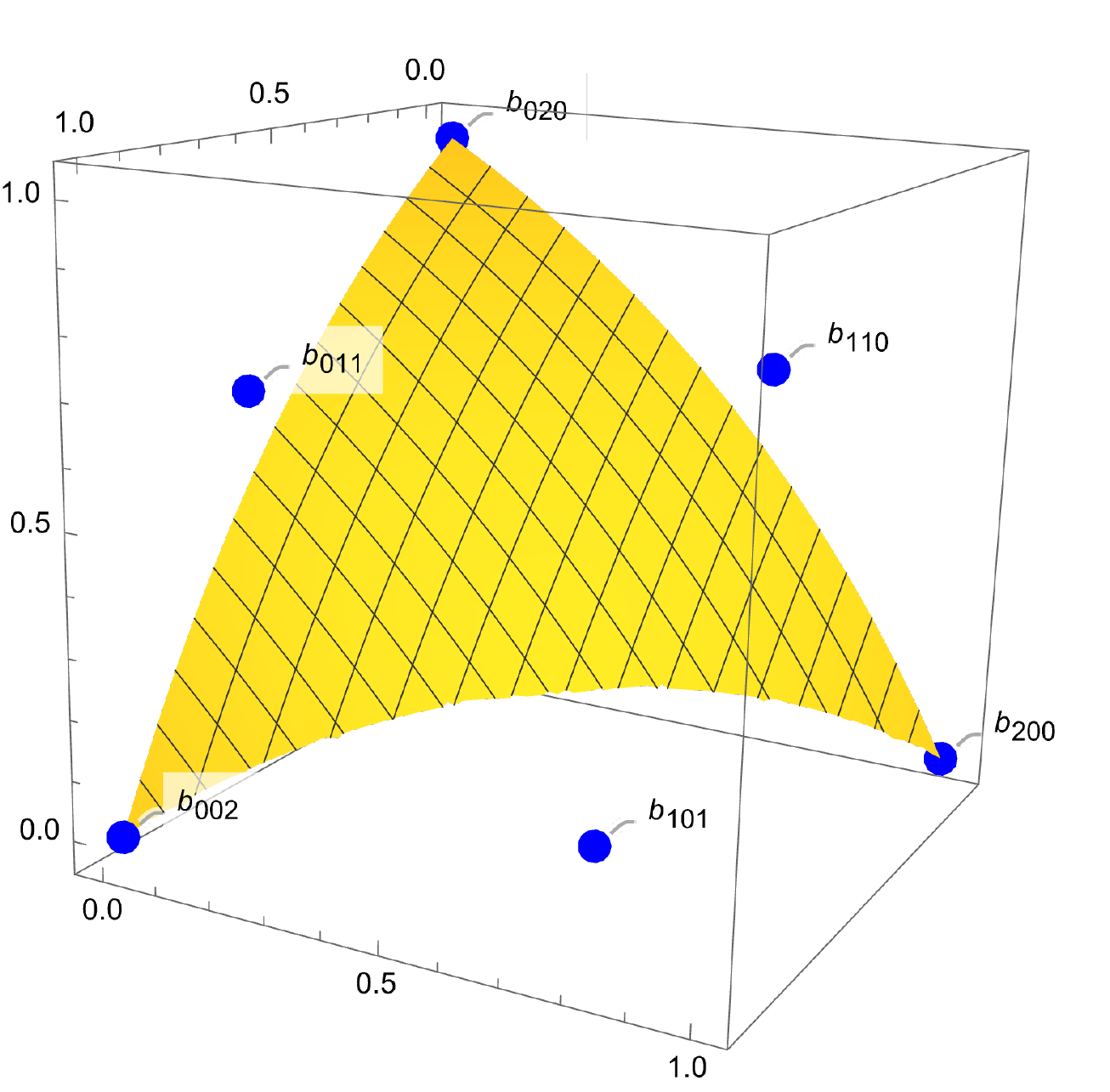}
    \includegraphics[height=2in]{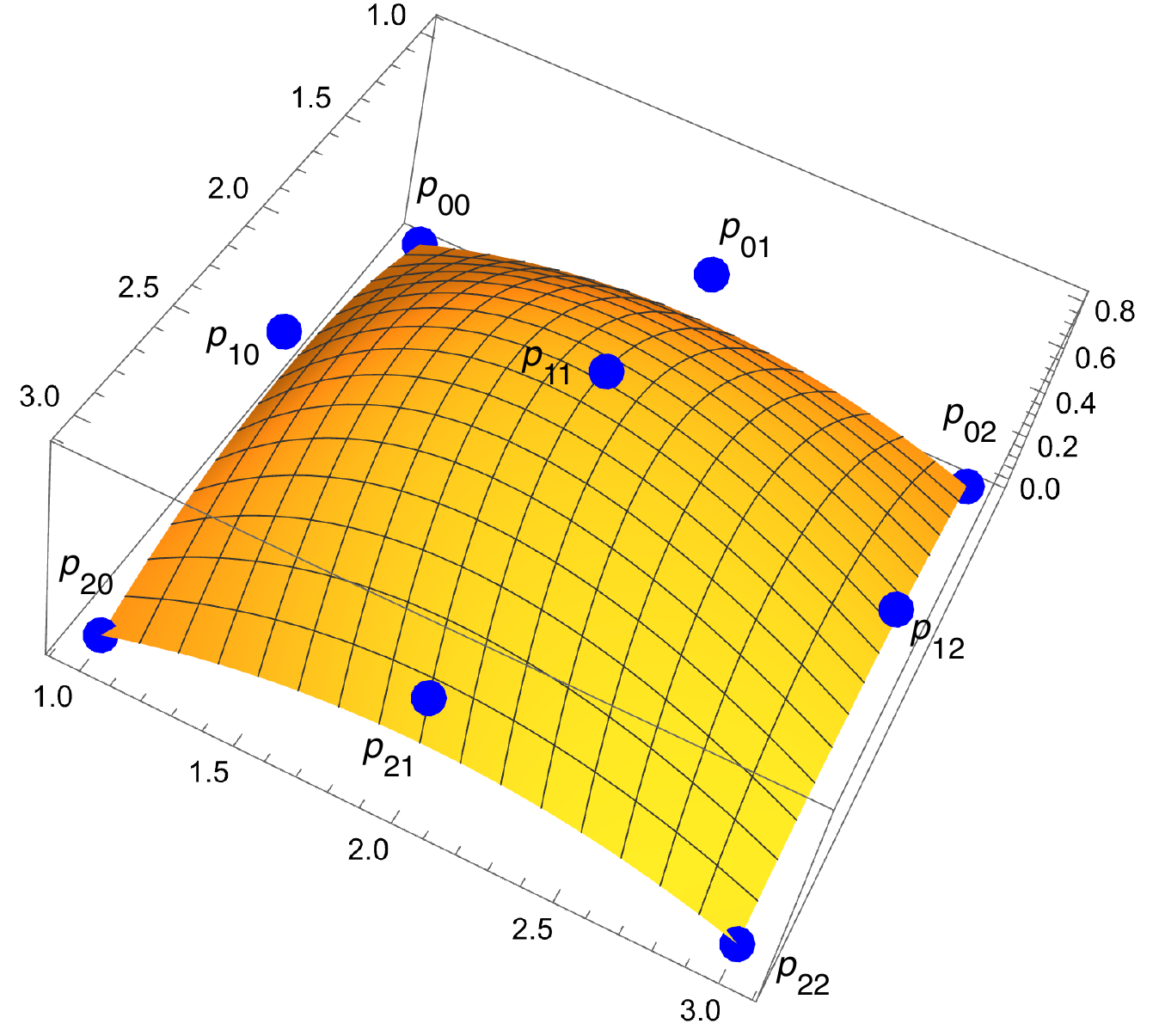}
	\caption{A biquadratic Bézier triangle and a biquadratic Bézier quadrilateral patch.}
	\label{fig:fig1}
\end{figure}
A biquadratic Bézier surface with a triangular input domain, typically called a Bézier triangle, is parametrized by six 3-dimensional control points, or 18 free parameters. These control points are labeled $b_{200}$, $b_{020}$, $b_{002}$, $b_{110}$, $b_{101}$, and $b_{011}$. $b_{200}$, $b_{020}$, and $b_{002}$ are the vertices of the triangle constituting the input domain. For any point $P$ on that triangle, the barycentric coordinates $(u,v)$ of that point are mapped to the surface by the $\Re^2 \rightarrow \Re^3$ function

\begin{equation}
	P(u,v)= b_{200}u^2 + b_{020}v^2+b_{002}w^2+2b_{110}uv+2b_{101}uw+2b_{011}vw
\end{equation}

From this function, we produce an implicit equation for each output dimension $(x,y,z)$ by subtracting $x$, $y$, and $z$ respectively, yielding

\begin{equation}
	f_{x}(u,v)= b_{200_{x}}u^2 + b_{020_{x}}v^2+b_{002_{x}}w^2+2b_{110_{x}}uv+2b_{101_{x}}uw+2b_{011_{x}}vw-x
 \end{equation}
\begin{equation}
	f_{y}(u,v)= b_{200_{y}}u^2 + b_{020_{y}}v^2+b_{002_{y}}w^2+2b_{110_{y}}uv+2b_{101_{y}}uw+2b_{011_{y}}vw-y
 \end{equation}
\begin{equation}
	f_{z}(u,v)= b_{200_{z}}u^2 + b_{020_{z}}v^2+b_{002_{z}}w^2+2b_{110_{z}}uv+2b_{101_{z}}uw+2b_{011_{z}}vw-z
\end{equation}

Since simultaneous satisfaction of these equations yields a valid point on the surface, the resultant of this system of 3 polynomials in $u$ and $v$ will be the implicit equation of the Bézier triangle. \citet{SEDERBERG198472} proposed implicitization of parametric polynomial surfaces using a method developed by \citet{dixon1909eliminant} which extends to three equations Cayley’s classic two polynomial resultant construction. 

By introducing two auxiliary variables  $\alpha$ and $\beta$, an augmented polynomial can be constructed

\begin{equation}
	\delta=  {\frac {1}{(u-\alpha)(v-\beta)}}  \text{det}
        \begin{bmatrix}
            f_{x}(u,v) & f_{y}(u,v) & f_{z}(u,v) \\
            f_{x}(u,\beta) & f_{y}(u,\beta) & f_{z}(u,\beta) \\
            f_{x}(\alpha,\beta) & f_{y}(\alpha,\beta) & f_{z}(\alpha,\beta) \\
        \end{bmatrix}
\end{equation}

that vanishes for any choice of $\alpha$ and $\beta$ if and only if $u$ and $v$ take values that satisfy the original implicit equations. This polynomial may be written as a product of three matrices

\begin{equation}
	\delta= 
        \begin{bmatrix}
            1 \\
            \alpha \\
            \beta \\
            ... \\
            \alpha\beta^3
        \end{bmatrix}
        \begin{bmatrix}
            ... & ... &  & ... & ... \\
            ... & ... &  & ... & ... \\
            ... & ... & ... & ... & ... \\
            ... & ... & & ... & ... \\
            ... & ... &  & ... & ...
        \end{bmatrix}
        \begin{bmatrix}
            1 \\
            u \\
            v \\
            ... \\
            u^3v
        \end{bmatrix}
\end{equation}

such that the entries of the center matrix (which we call the Cayley matrix, of maximum order 2*2*2 = 8 for biquadratic polynomials) are the coefficients (which will be linear polynomials in $x$, $y$, and $z$) for each non-zero coefficient combination of $u$, $v$, $\alpha$, and $\beta$. The determinant of the Cayley matrix, if it is not identically zero \footnote{certain polynomial systems may require a more complex augmentation, such as a row with terms like $\lambda f_x(u,\beta)+ \mu f_x(\alpha,v)$ to avoid a zero determinant, but they are not needed here.}
, is then the resultant, termed the Dixon resultant \citep{10.1145/190347.190372}, of the three initial implicit equations; however, the order of the determinant may render this calculation impractical.

This approach may also be used to calculate the implicit equation of a biquadratic Bézier surface with a square input domain, termed a \emph{patch}. This surface is parametrized by a 3x3 grid of nine control points labeled $p_{ij}$ where $i$ and $j$ are grid indices:
\begin{equation}
	P(u,v)= \sum _{i=0}^2 \sum _{j=0}^2 {\frac {2}{i!(2-i)!}} u^i(1-u)^{2-i} {\frac {2}{j!(2-j)!}} v^j(1-v)^{2-j} p_{ij}
\end{equation}

for any point $(u,v)$ on the unit square.

Following the same approach outlined above for this mapping yields the three implicit sub-equations

\begin{equation}
    \begin{aligned}
    	0&= p_{00_{x}}(1-u)^2(1-v)^2 + 2p_{01_{x}}v(1-u)^2(1-v) + p_{02_{x}}v^2(1-u)^2 + 2p_{10_{x}}u(1-u)(1-v)^2 \\
         & + 4p_{11_{x}}uv(1-u)(1-v) + 2p_{12_{x}}uv^2(1-u) + p_{20_{x}}u^2(1-v)^2 + 2p_{21_{x}}u^2v(1-v) + p_{22_{x}}u^2v^2 - x
    \end{aligned}
\end{equation}
\begin{equation}
    \begin{aligned}
    	0&= p_{00_{y}}(1-u)^2(1-v)^2 + 2p_{01_{y}}v(1-u)^2(1-v) + p_{02_{y}}v^2(1-u)^2 + 2p_{10_{y}}u(1-u)(1-v)^2 \\
         & + 4p_{11_{y}}uv(1-u)(1-v) + 2p_{12_{y}}uv^2(1-u) + p_{20_{y}}u^2(1-v)^2 + 2p_{21_{y}}u^2v(1-v) + p_{22_{y}}u^2v^2 - y
    \end{aligned}
\end{equation}
\begin{equation}
    \begin{aligned}
    	0&= p_{00_{z}}(1-u)^2(1-v)^2 + 2p_{01_{z}}v(1-u)^2(1-v) + p_{02_{z}}v^2(1-u)^2 + 2p_{10_{z}}u(1-u)(1-v)^2 \\
         & + 4p_{11_{z}}uv(1-u)(1-v) + 2p_{12_{z}}uv^2(1-u) + p_{20_{z}}u^2(1-v)^2 + 2p_{21_{z}}u^2v(1-v) + p_{22_{z}}u^2v^2 - z
    \end{aligned}
\end{equation}

The resultant of these equations may again be calculated using Dixon's approach.

\section{Results}

We used the symbolic computation library SymPy \citep{10.7717/peerj-cs.103} to construct Cayley matrices for the general biquadratic Bézier triangle and quadrilateral using their parametric implicit equations. Because SymPy is slow at calculating large symbolic determinants, we used Mathematica to calculate the determinant expansions, yielding the implicit equations of the general biquadratic Bézier triangle and quadrilateral. The Cayley matrix for the biquadratic triangle is 5x5, since the terms $u^3$, $u^3v$ and $u^2v$ do not appear in the $\delta$ polynomial, and $\delta$ is of the form

\begin{equation}
	\delta= 
        \begin{bmatrix}
            1 \\
            \alpha \\
            \beta \\
            \beta^2 \\
            \alpha\beta
        \end{bmatrix}
        \begin{bmatrix}
            m_{00} & m_{10} & m_{20} & m_{30} & m_{40} \\
            m_{01} & m_{11} & m_{21} & m_{31} & m_{41} \\
            m_{02} & m_{12} & m_{22} & m_{32} & m_{42} \\
            m_{03} & m_{13} & m_{23} & m_{33} & 0 \\
            m_{04} & m_{14} & 0 & m_{34} & m_{44}
        \end{bmatrix}
        \begin{bmatrix}
            1 \\
            u \\
            v \\
            u^2 \\
            uv
        \end{bmatrix}
\end{equation}

where $m_{ij}=A_{ij}x+B_{ij}y+C_{ij}z + D_{ij}$ and $A,B,C,D$ are multilinear polynomials of control point components $b_{***}$.

Using the same construction, we produce a Cayley matrix for the biquadratic quadrilateral, which is 8x8 since all terms $u^{(0...3)}v^{(0..1)}$ appear in the polynomial. 

The implicit equations are too long (353 kB of plaintext for the triangle and 430 MB for the quadrilateral) to be printed here, but they are provided as text files and as standalone Python functions at \href{https://github.com/focolab/biquadratic-implicitization}{https://github.com/focolab/biquadratic-implicitization}. The implicit equation for the triangle consists of terms up to degree 5 in $x$,$y$, and $z$ and requires 12,986 multiplications and 269 additions, and the implicit equation for the quadrilateral consists of terms up to degree 8 in $x$, $y$, and $z$ and requires 16,621,762 multiplications and 138,561 additions. While these equations are exact, they consist of sums of many product terms and may therefore susceptible to round-off or numerical overflow error when evaluated using finite precision variables.

Because our implicit equation for the Bézier quadrilateral is much larger than our implicit equation for the Bézier triangle, evaluations for the quadrilateral are much slower (2.983s per evaluation during our testing) than for the triangle (0.0024s per evaluation). However, we can approximate evaluation by computing the numerical determinant of the Cayley matrix, and such approximations can be completed relatively quickly (0.0024s per evaluation for the quadrilateral and 0.0003s per evaluation for the triangle in our testing). Though both the implicit function and the numerical determinant computation may be prone to round-off error in certain cases, for example when inter-control point distances are very small relative to overall control point magnitudes (Figure 2), informal testing suggests numerically evaluated Cayley matrix determinants may adequately approximate evaluation of the full implicit functions in typical scenarios.

\subsection{Coordinate transformation reduces computations}
Without loss of generality, equivalent implicit equations of lower complexity may be produced for both the triangle and quadrilateral by applying a coordinate  transformation to the set of control points such that control point components are fixed to zero or one, prior to computing the resultant.  By mapping points $b_{200}$, $b_{020}$, and $b_{002}$ for the triangle and $p_{00}$, $p_{02}$, and $p_{20}$ for the quadrilateral to the points $(0,0,0)$ and $(1,0,0)$ and to the plane $z=0$ respectively, amounting to a translation, a rotation, and uniform scaling, we zero out 6 of 18 coefficients (and eliminate 7 degrees of freedom) from the system of parametric equations used to compute the Dixon resultant. This transformation is given by: 

\begin{equation}
    f(\vec{p})= \textbf{R}_{2}(g(\vec{p}) - g(\vec{c_{1}})), \qquad
    g(\vec{q})= \textbf{R}_{1} {\frac{\vec{q}}{\lVert\vec{c_{1}}-\vec{c_{2}}\rVert}}
\end{equation}

where $\vec{c_{1}}$, $\vec{c_{2}}$, and $\vec{c_{3}}$ are the control points $b_{200}$, $b_{020}$, and $b_{002}$ respectively of a Bézier triangle or $p_{00}$, $p_{02}$, and $p_{20}$ of a Bézier quadrilateral; $\textbf{R}_{1}$ is a rotation matrix which will align the normal $(\vec{c_{1}}-\vec{c_{2}})\times(\vec{c_{3}}-\vec{c_{2}})$ with the vector $(0,0,1)$; and $\textbf{R}_{2}$ is a rotation matrix around the $z$-axis which will place $(g(\vec{c_{2}}) - g(\vec{c_{1}}))$ on the $x$-axis.

Using this coordinate transformation trick, the number of multiplications required for evaluation of the implicit function is reduced from 12,986 to 5,278 for the triangle and from 16,621,762 to 4,719,155 for the quadrilateral, and the number of required additions remains unchanged. Additionally, because the magnitude of control point components, and consequently their intermediate monomial products, are rescaled to be near unity order of magnitude (aside from pathologically thin triangles and patches), the coordinate transformation approach will be more robust to finite-precision effects (Figure 2).

\begin{figure}%
    \centering
    \centerline{\includegraphics[height=4in]{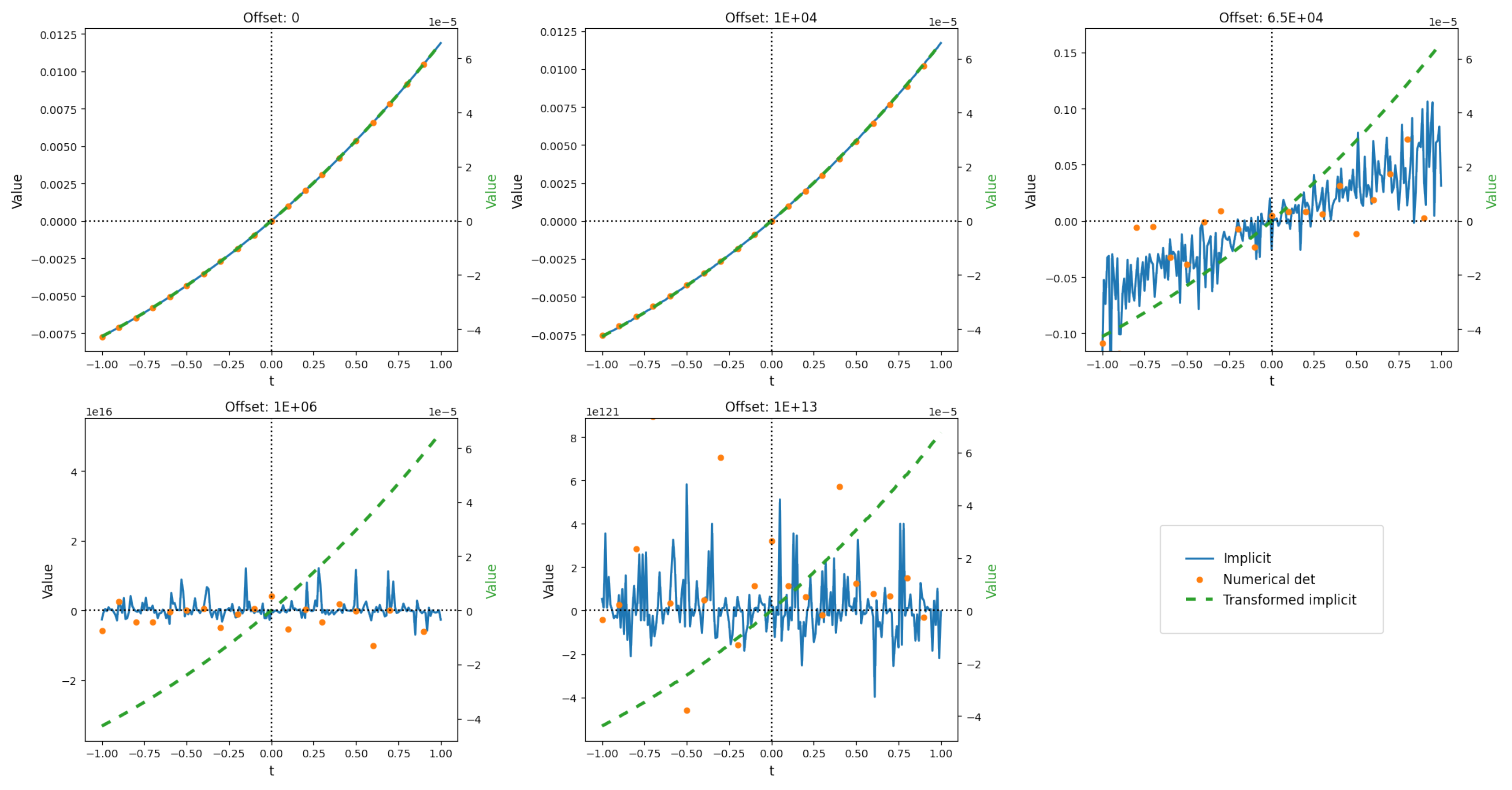}}
    \caption{Some finite-precision failure scenarios can be avoided by use of coordinate pre-transformation.}%
    \medskip
    \small
    \justifying
    Limited-precision representations can cause error in the implicit function for the biquadratic Bézier triangle and its numerical determinant approximation. We construct five input cases by adding offsets of magnitudes $\{0,1E4,6.5E4,1E6,1E13\}$ to the control points of the triangle in Figure 1. For each case, we identify a point $\vec{p}$ on the triangle surface $(u=.33,v=.33)$. We then calculate a unit vector $\hat{v}={\frac{\vec{p}}{\lVert\vec{p}\rVert}}$ and evaluate the implicit function (blue) and numerical determinant approximation (orange dots) along the line segment $\vec{p}+t\hat{v}, -1 \leq t \leq 1$, which should have a zero crossing at $t=0$. Initial control points are $b_{200}=(1,0,0)$, $b_{020}=(0,1,0)$, $b_{002}=(0,0,1)$, $b_{110}=(.65,.65,0)$, $b_{101}=(.65,0,.65)$, and $b_{011}=(0,.65,.65)$, and evaluations are carried out in Python using 64-bit floats.
    \label{fig:fig2}%
\end{figure}

\section{Discussion}

We have generated general implicit equations for biquadratic Bézier triangles and quadrilaterals, and we have shown that these equations can be used to determine the relationship (on, inside, or outside) to a given biquadratic Bézier surface for an arbitrary point in space. Biquadratic surfaces can express a variety of smoothly curved surfaces, but they do not offer enough degrees of freedom to be easily stitched together into arbitrary composite surfaces where $C^1$ continuity is preserved across seams. Consequently, flexible object modeling systems commonly use bicubic surfaces instead. In principle, the method used here to calculate implicit equations for the biquadratic case may be applied to higher order Bézier surfaces, but for general cases, the number of terms of the resultant polynomial explodes and exceeds practical computational constraints. In the absence of a closed-form symbolic function, numerical evaluation of the determinants of higher order Cayley matrices may be a satisfactory approximation of their implicit functions for many applications.  These methods should map extremely well to GPU architectures since they require no branching or iteration.

\section{Acknowledgements}

J.B. and S.K. contributed to conception, theory, implementation, and writing. This work was funded by the Weill Institute for Neurosciences Neurocomputing Grant, the Weill Neurohub Pillars Program, and the NIH National Institute of General Medical Sciences (R35GM124735).

\bibliographystyle{unsrtnat}
\bibliography{references}

\end{document}